\documentclass[prl,twocolumn,showpacs,preprintnumbers,amsmath]{revtex4}
\usepackage{color}
\newcommand{\be}{\begin{equation}}
\newcommand{\ee}{\end{equation}}

\newcommand{\bse}{\begin{subequations}}
\newcommand{\ese}{\end{subequations}}
\newcommand{\bea}{\begin{eqnarray}}
\newcommand{\eea}{\end{eqnarray}}
\newcommand{\ba}{\begin{array}}
\newcommand{\ea}{\end{array}}
\newcommand{\vect}{\mathbf}
\def \th {\theta^{\mu\nu}}

\def\P{Poincar\'e }

\def\nc{noncommutative\ }
\def\ncy{noncommutativity\ }

\makeatother

\begin{document}
\title{{ A  Realization of the Cohen-Glashow Very Special Relativity}}%
\author{{M. M. Sheikh-Jabbari$^1$ and A. Tureanu$^2$}}
\affiliation{$^1$School of Physics, Institute for Research in
Fundamental Sciences (IPM), P.O.Box 19395-5531, Tehran, Iran}
\email{jabbari@theory.ipm.ac.ir}
\affiliation {$^2$Department of Physics, University of Helsinki and
Helsinki Institute of Physics, P.O.Box 64, FIN-00014 Helsinki,
Finland} \email{anca.tureanu@helsinki.fi}

\begin{abstract}
We show that the Cohen-Glashow Very Special Relativity (VSR) theory
\cite{VSR} can be realized as the part of the Poincar\'e symmetry
preserved on a noncommutative Moyal plane with light-like
noncommutativity. Moreover, we show that the three subgroups
relevant to  VSR can also be realized in the noncommutative
space-time setting. For all these three cases the noncommutativity
parameter $\th$ should be light-like ($\th\theta_{\mu\nu}=0$). A
fixed constant noncommutativity parameter respects the $T(2)$
subgroup of Lorentz, while for the $E(2)$ and $SIM(2)$ cases the
form of noncommutativity among the coordinates should be of linear
(Lie algebra) and quadratic (quantum group) type, respectively. We
discuss some physical implications of this realization of the
Cohen-Glashow VSR.

\begin{center}
(to appear in Phys. Rev. Lett.)
\end{center}

\end{abstract}

\maketitle


The Special Theory of Relativity postulates the \P group as the
symmetry of Nature. It is believed, however, that at very high
energies, the usual description of space-time in terms of a smooth
manifold would break down, and together with it, the Lorentz
invariance of physical theories. Various possible departures from
Lorentz invariance at high energies have been studied,  both
theoretically and experimentally (see \cite{VSR,CG,CK} and
references therein). The problem addressed in \cite{VSR} is whether
Lorentz invariant theories, like the Standard Model, could emerge as
effective theories from a more fundamental scheme, perhaps operative
at the Planck scale, which is invariant under Very Special
Relativity groups, but not invariant under the full \P group.

Very Special Relativity (VSR) has been introduced in Ref. \cite{VSR}
as symmetry under certain subgroups of \P group, which contain
space-time translations and at least a 2-parametric proper subgroup
of the Lorentz transformations, isomorphic to that generated by
$K_x+J_y$ and $K_y-J_x$, where $\vect J$ and $\vect K$ are the
generators of rotations and boosts, respectively. These subgroups of
the Lorentz group share the remarkable property that, when
supplemented with $T$, $P$ or $CP$, they will be enlarged to the
full Lorentz group. This can be taken as definition of VSR.

The requirement of energy-momentum conservation should be preserved
in VSR, consequently in all its realizations the translational
symmetry should be contained. Besides generators of translations
$P_\mu$, the minimal version of VSR includes the Abelian subgroup of
the Lorentz group, $T(2)$, generated by $T_1=K_x+J_y$ and
$T_2=K_y-J_x$. The group $T(2)$ can be identified with the
translation group on a two-dimensional plane. The other larger
versions of VSR are obtained by adding one or two Lorentz generators
to $T(2)$, which have
geometric realization on the two-dimensional plane:\\
{\it i)} $E(2)$, the group of two-dimensional Euclidean motion,
generated by $T_1,\ T_2$ and $J_z$, with the structure:
\be [T_1,T_2]=0,\ \ [J_z,T_1]=-iT_2,\ \ [J_z,T_2]=iT_1;\ee
{\it ii)} $HOM(2)$, the group of orientation-preserving similarity
transformations, or homotheties, generated by $T_1,\ T_2$ and $K_z$,
with the structure
\be [T_1,T_2]=0,\ \ [T_1,K_z]=iT_1,\ \ [T_2,K_z]=iT_2;\ee
{\it iii)} $SIM(2)$, the group isomorphic to the four-parametric
similitude group, generated by $T_1,\ T_2,\ J_z$ and $K_z$.

When attempting to construct a concrete realization of the VSR
symmetry as a fundamental scheme within a ``master theory'', most
certainly nonlocal \cite{VSR}, and which would lead in the
low-energy limit to an effective \P invariant theory, one runs into
the problem of the representation content of the ``master theory".
The Lorentz subgroups involved in VSR have only one-dimensional
representations, unlike the Lorentz group. The representations of
$T(2), E(2), HOM(2)$ and $SIM(2)$ are automatically representations
of the Lorentz group, but the reciprocal is not valid. As a result,
if we construct the ``master theory" based on the one-dimensional
representations of the VSR subgroups, when requiring also $P$, $T$
or $CP$ invariance, although the theory becomes invariant under the
whole Lorentz group, the one-dimensional representations will not
change. As a result, the effective theory would be doomed by its
very poor representation content. Another possibility is to use in
the realization of VSR the representations of the full Lorentz
group, but add a Lorentz-violating factor, such that the symmetry of
the Lagrangian is reduced to one of the VSR subgroups of the Lorentz
group. However, such an approach can hardly provide a {\it
fundamental theory}, given that its symmetry does not match its
representation content.

This contradiction can be resolved if we abandon the reasoning in
terms of Lie groups/algebras and extend the discussion to (deformed)
Hopf algebras. In the framework of Hopf algebras, there exist
deformations which leave the structure of the algebra (commutation
relation of the generators) untouched, but affect other properties
of the Hopf algebra, i.e. the co-algebra structure
\cite{monographs}. Since the commutation relations of generators are
not deformed, it follows automatically that the Casimir operators
are the same and the representation content of the deformed Hopf
algebra is identical to the one of the undeformed algebra. On the
other hand, the deformation of the co-algebra structure reduces the
symmetry of the scheme. Such deformations are the twists introduced
in Ref. \cite{Drinfeld83}, which turned out to provide a powerful
concept, facilitating the systematic approach to deformation
quantization \cite{monographs}.

Noncommutative quantum field theories (NC QFTs) are field theories
constructed on space-times whose coordinates satisfy the commutation
relations
\be\label{cr}%
[x^\mu,x^\nu]=i\th\,,
\ee
where $\theta$ can be a function of coordinates (with the condition
that it satisfies the Jacobi identity). The commutation relations
\eqref{cr} usually spoil the Lorentz invariance (and sometimes also
the translational invariance) of the NC QFTs; however, these
theories remain invariant under the subgroup of the \P group which
preserves the covariance of \eqref{cr} (see \cite{LAG} for the case
of constant $\theta$).

The essential element for our discussion is that NC QFTs possess
symmetry under various twisted \P algebras, depending on the
structure of $\theta$ \cite{CKNT,CPrT} (see also \cite{LW}). The
advantage of using the twisted \P language for constructing physical
theories is that, in spite of the lack of full Lorentz symmetry, the
fields carry representations of the full Lorentz group
 \cite{CKTZZ,CNST} and the spin-statistics relation is still valid
 ; the deformation then
appears in the product of the fields (interaction terms).

Although we reviewed mainly the technical merits of NC QFT as a
candidate for realizing VSR, we should recall the main motivation
for introducing the NC space-time, based on the interplay of quantum
theory and classical gravity \cite{DFR}, as well as the emergence of
NC QFTs as low-energy effective theories from string theory in
Kalb-Ramond background field \cite{AAS,SW}.

In this letter we show that the NC spaces with light-like \ncy
\cite{AGM} offer a natural setting which realizes VSR, hence
providing us with the well-studied theoretical setting of NC QFT,
with twisted \P symmetry, as physical framework for VSR.

{\it Space-Time Noncommutativity and VSR.}---To start with,  we
focus on the NC spaces defined through \eqref{cr}. Since $\th$ is an
anti-symmetric two tensor, NC spaces can be classified according to
the two Lorentz invariants
\be\Lambda^4\equiv \theta_{\mu\nu}\theta^{\mu\nu},\ \ \
   L^4\equiv
   \epsilon^{\alpha\beta\mu\nu}\theta_{\mu\nu}\theta_{\alpha\beta}.
\ee%
$\Lambda^4$ is related to the \ncy scale, the scale where
noncommutativity effects will become important, while $L^4$ is
related to the smallest (space-time) volume that we can measure in a
\nc theory.

Depending on whether $L^4$ and $\Lambda^4$ are positive, zero or
negative one can recognize nine cases. The $L^4\neq 0$ cases cannot
be obtained as a decoupling (low energy) limit of open string theory
and do not lead to a unitary NC QFT theory \cite{AGM} and hence are
not usually considered. (However, the $\Lambda ^4=0$, $L^4\neq 0$
case is the famous Doplicher-Fredenhagen-Roberts \cite{DFR}
noncommutative space.)

 For $L^4=0$,
depending on the value of $\Lambda^4$ there are three types of
noncommutative spaces:\\ {\it i)} $\Lambda^4>0$ -- space-like
(space-space) noncommutativity;\\ {\it ii)} $\Lambda^4<0$ --
time-like (time-space) noncommutativity;\\ {\it iii)} $\Lambda^4=0$
-- light-like
noncommutativity.\\
When $\Lambda$ is  constant, for the case {\it ii)}, it has been
shown that there is no well-defined decoupled field theory limit for
the corresponding open string theory \cite{AGM}. In the field theory
language this shows itself as instability of the vacuum state and
non-unitarity of the field theory on time-like NC space \cite{GM}.
For the space-like case {\it i)} and light-like case {\it
iii)}, noncommutative field theory limits are well-defined and the
corresponding field theories are perturbatively unitary.

Depending on the structure of the r.h.s. of \eqref{cr}, there exist
three types of NC deformations of the space-time which can be
realized through
twists of the \P algebra \cite{CKNT,CPrT,LW}:\\
1) \textit{Constant $\th$}: the Heisenberg-type commutation
relations, defining the \textit{Moyal space}:
\be\label{heisenberg} [x^\mu,x^\nu]=i\th \,, %
\ee
where $\th$ is a constant antisymmetric matrix. \\
2) \textit{Linear $\th$}, with the Lie-algebra type commutators:%
\be\label{lie}%
 [x^\mu,x^\nu]=iC_\rho^{\mu\nu} x^\rho \,, %
\ee%
describing an (associative but) noncommutative space if
$C_\rho^{\mu\nu}$ are structure constants of an associative Lie
algebra.\\ %
3) \textit{Quadratic noncommutativity}, the quantum group type of
commutation relations:
\be\label{qg}%
[x^\mu,x^\nu]=\frac{1}{q}R_{\rho\sigma}^{\mu\nu} x^\rho x^\sigma \,. %
\ee

All the above-mentioned cases of noncommutative space-time have
originally been studied in \cite{CDP} with respect to the
formulation of NC QFTs on those spaces. Only in the case 1) the
translational invariance is preserved in all the directions of
space-time.
Since the translation symmetry is one of the requirements of the
Cohen-Glashow VSR \cite{VSR}, only the Moyal NC space-time is
relevant to the VSR theories and therefore here we mainly focus on
the Moyal case. We shall briefly discuss the linear and quadratic
$\th$ cases, since in special conditions, all these types of
noncommutativity can be put in a relation to certain Lorentz
subgroups relevant to VSR.

{\it $T(2)$ symmetry as light-like noncommutativity}.---Motivated by
the above arguments, we set about finding a configuration of the
antisymmetric matrix $\th$ which would be invariant under the $T(2)$
subgroup of the Lorentz group - the only of the VSR subgroups which
admits invariant tensors, as also noted in \cite{VSR}. If we denote
the elements of the $T(2)$ subgroup by
\be\label{l1l2} \Lambda_1=e^{i\alpha T_1}\ \ \ \mbox{and}\ \ \
\Lambda_2=e^{i\beta T_2}, \ee
the invariance condition for the tensor $\theta^{\mu\nu}$ is
written as:
\be\Lambda^{\phantom{i}\mu
}_{i\phantom{\mu}\alpha}\Lambda^{\phantom{i}\nu
}_{i\phantom{\nu}\beta}\theta^{\alpha\beta}=\theta^{\mu\nu},\ \ \
i=1,2,\ee
and infinitesimally:
\be\label{inv_eq} T^{\phantom{i}\mu
}_{i\phantom{\mu}\alpha}\theta^{\alpha\nu}+T^{\phantom{i}\nu
}_{i\phantom{\nu}\beta}\theta^{\mu\beta}=0,\ \ \ i=1,2.\ee
The nonvanishing elements of the matrix realizations of the
generators $T_1$ and $T_2$ are (see, e.g., the monograph
\cite{CH}):
$(T_1)^0_{\phantom{0}1}=(T_1)^1_{\phantom{1}0}=(T_1)^1_{\phantom{1}3}=-(T_1)^3_{\phantom{3}1}=i$
and
$(T_2)^0_{\phantom{0}2}=(T_2)^2_{\phantom{2}0}=(T_2)^2_{\phantom{2}3}=-(T_2)^3_{\phantom{3}2}=i$.

Plugging these values into \eqref{inv_eq}
we find the solution%
\be\label{ll-ncy}%
 \theta^{0i}=-\theta^{3i},\ \ \ i=1,2,%
\ee%
all the other components of the antisymmetric matrix
$\theta^{\mu\nu}$ being zero. Note that to obtain the above result
we did not assume any special form for the $x$-dependence of $\th$
and hence this holds for either of the three constant, linear and
quadratic cases.

With the above $\th$ we see that $ \Lambda^4=L^4=0$, that is {\it
a light-like $\theta^{\mu\nu}$ is invariant under $T(2)$}.

One may use the light-cone frame coordinates%
\be\label{light-cone-coor}%
x^\pm=(t\pm x^3)/2,\qquad x^i\,, \ i=1,2.%
\ee%
In the above coordinate system the only non-zero components of the
light-like \ncy \eqref{ll-ncy} is
$\theta^{-i}=\theta^{0i}=-\theta^{3i}$ (and
$\theta^{+-}=\theta^{+i}=\theta^{ij}=0$). In the light-cone
coordinates (or light-cone gauge) one can take $x^+$ to be the
light-cone time and $x^-$ the light-cone space direction. In this
frame, (light-cone) time commutes with the space coordinates. In the
light-cone $(+,-,1,2)$ basis
\begin{eqnarray}\label{theta}
\theta^{\mu\nu}=\left(
\begin{array}{cccc}
0 & 0 & 0  & 0 \\
0 & 0 & \theta  & \theta' \\
0& -\theta & 0  & 0 \\
0 & -\theta'  & 0 & 0
\end{array}
\right).
\end{eqnarray}

{\it $E(2)$ and $SIM(2)$ invariant NC spaces}.---A constant
$\theta^{-i}$ breaks rotational invariance in the $(x^1,x^2)$-plane
and hence larger VSR subgroups are not possible in the Moyal NC
space case. The $E(2)$ invariant case can be realized in the linear,
Lie-algebra type noncommutative spaces and $SIM(2)$ can be realized
by quadratic noncommutativity.
\paragraph{The $E(2)$ case:}
Recalling that $J_z$ is the generator of rotations in the
$(x^1,x^2)$-plane while keeping $x^\pm$ invariant, and that $x^i$
are invariant under $K_z$, which acts as scaling on $x^-$ and $x^+$,
the $E(2)$ case is realized when $\theta^{-i}$ is proportional to
$x^i$ and is independent of $x^-$. Noting that the only invariant
tensors under $J_z$ are $\delta_{ij}$ and $\epsilon_{ij}$, then
$\theta^{-i}$ has to be proportional to the product of either of
these invariant tensors with the only available vector on the
$(x^1,x^2)$-plane:
\bse\label{E2-theta}%
\begin{align}%
[x^-, x^i] &=i\ell \epsilon^{\,ij} x^j, \quad \mbox{or}\ \\ [x^-,
x^i] &=i\ell x^i.
\end{align}
\ese%
With the above choices, the translational symmetry along $x^\pm$ is
preserved while along $x^i$ it is clearly lost.

Instead of $x^i$ coordinates we may work with the cylindrical
coordinates on $x^-, x^1,x^2$ space, with the axis of the cylinder
along $x^-$. If we denote the radial and angular coordinate on the
$(x^1,x^2)$-plane by $\rho$ and $\phi$, the case (\ref{E2-theta}a)
is then described by:
\be%
[x^-,\rho]=0,\quad [\rho, e^{\pm i\phi}]=0, \quad [x^-, e^{\pm
i\phi}]=\pm \lambda e^{\pm i\phi}.%
\ee%
This space is a collection of NC cylinders of various radii. There
is a twisted \P algebra which provides the symmetry for this case
while  the other case cannot be generated by a twist
\cite{progress}. In the above $\ell$ and $\lambda$ are deformation
parameters of dimension length.
\paragraph{The $SIM(2)$ case:}
From the above discussions it becomes clear that to have both the
$K_z$ and $J_z$ invariant noncommutative structures we should take
$\theta^{-i}$ linear in both $x^-$ and $x^i$, therefore the two
possibilities are
\bse\label{SIM2-theta}%
\begin{align}%
[x^-, x^i] &=i\frac{\xi-1}{\xi+1}\ \epsilon^{\,ij} \{x^-, x^j\},\ \quad \mbox{or}\ \\
[x^-, x^i] &=i\tan\xi \{x^-, x^i\},
\end{align}
\ese%
preserving translational symmetry only along $x^+$ (where $\xi$ is
the dimensionless deformation parameter).

For neither of the above cases there is any twisted \P algebra of
the form discussed in \cite{LW} which provides these commutators
\cite{progress}. The case (\ref{SIM2-theta}b) in the above mentioned
cylindrical coordinates $x^-, \rho, \phi$ takes the familiar form of
a quantum (Manin) plane.

As mentioned above, the Cohen-Glashow VSR requires translation
invariance, which is only realized in the constant $\th$ case,
therefore we continue with the discussion of QFTs on the
light-like Moyal plane, as the VSR-invariant theories. Further
analysis of the linear and quadratic cases will be postponed to
future works \cite{progress}.

{\it NC QFT on Light-like Moyal plane as VSR invariant theory}.---So
far we have shown that a Moyal plane with light-like
noncommutativity is invariant under the $T(2)$ VSR. Consequently,
the NC QFT constructed on this space possess also the same symmetry
\cite{LAG}, as well as twisted \P symmetry \cite{CKNT,CPrT}. For any
given QFT on commutative Minkowski space its noncommutative
counterpart, NC QFT, is  obtained by replacing the usual product of
functions (fields) with the nonlocal Moyal $*$-product (for a review
on NC QFTs, see \cite{NCQFT-review}):
\be\label{Moyal-product-def}%
(\phi*\psi)(x)=\phi(x)\
e^{\frac{i}{2}\th\overleftarrow{\partial_\mu}\overrightarrow{\partial_\nu}}\
\psi(x)\ . %
\ee%
Due to the twisted \P symmetry, the fields carry representations of
the full Lorentz group \cite{CKTZZ,CNST},  but they admit
transformations only under the stability group of light-like $\th$,
$T(2)$.

NC QFTs are $CPT$-invariant and satisfy the spin-statistics relation
\cite{CPT,CNT,spin-stat}. However, as shown in \cite{CPT} for NC
QED, $C$, $P$ and $T$ symmetries are not individually preserved: for
the time-space noncommutativity, which comprises also the light-like
case, $P$ invariance requires also the transformation
$\theta^{0i}\to-\theta^{0i}$, $C$ invariance requires $\th\to-\th$
and $T$-invariance requires $\theta^{ij}\to-\theta^{ij}$. Since
$\th$ is invariant on the Moyal space, these transformations cannot
occur. Consequently, requiring $P$, $T$ or $CP$ invariance from NC
QFT with $T(2)$ VSR symmetry is equivalent to taking $\theta\to 0$,
in which case the emerging theory has full \P symmetry, as predicted
in \cite{VSR}.

{\it Discussion and Outlook}.---We have shown that light-like Moyal
NC space provides a consistent framework for $T(2)$ Cohen-Glashow
VSR-invariant theories. The other VSR groups, $E(2)$, $SIM(2)$ and
$HOM(2)$ are ruled out, if  the origin of Lorentz violation is in
the NC structure of space-time, since the corresponding NC spaces
are not translationally invariant. The realization of VSR as NC
theories has several advantages: 1) Despite the lack of full Lorentz
symmetry, one can still label fields by the Lorentz representations.
For the NC QFTs we can rely on the basic notions of fermions and
bosons, spin-statistics relation and CPT theorem
\cite{CPT,CNT,spin-stat}; 2) There is a simple recipe for
constructing the NC version of any given QFT. Noncommutativity
introduces a structure, fixing the form of the VSR-invariant action;
3) In the NC setting we only deal with a single deformation
parameter (the coordinates on $(x^1,x^2)$-plane can be chosen such
that $\theta'$ in \eqref{theta} is zero).

The parameter $\theta$ of the NC QFT realization of $T(2)$ VSR is of
dimension length-square and it defines the noncommutativity scale
$\Lambda_{NC}=1/\sqrt \theta$. To find bounds on $\Lambda_{NC}$ we
need to compare results based on the NC models to the existing
observations and data. These data can range from atomic spectroscopy
and Lamb-shift (see, e.g., \cite{Lamb-shift}) to particle physics
bounds on the electric-dipole moments of elementary particles. Since
the structure of the terms involving the light-like NC parameter is
essentially the same as in the more-studied case with space-space
noncommutativity (as it stems from the same *-product
\eqref{Moyal-product-def}), and based on explicit calculations
\cite{progress}, we can infer that the bounds on the NC parameter
will be of the same order of magnitude as the previously calculated
ones, i.e. $\Lambda_{NC}>10\ \mbox{TeV}$ \cite{Lamb-shift} (see also
\cite{Kostelecky} for a similar bound coming from clock-comparison
experiments).

To construct a particle physics model based on the NC realization
for VSR we need to fix a light-like $\th$ in any of the
noncommutative models constructed so far, with generic
noncommutativity. Although various basic features are common to all
NC QFTs constructed with the $*$-product \eqref{Moyal-product-def},
the light-like case has some specific features which are not shared
by other NC QFTs on Moyal space-time. Moreover, the light-like NC
QFTs are also unitary field theories \cite{AGM}. The construction of
consistent light-like NC models may lead to the derivation of more
stringent bounds on $\Lambda_ {NC}$. The light-like NC QFTs are
expected to have many features in common with the space-space NC
case, which has been thoroughly studied in the literature, and main
results of that case should also hold for the light-like case.
Clarifying which of the results carry over to the light-like NC QFT
will be discussed in a forthcoming communication \cite{progress}.

We warmly thank Masud Chaichian for many valuable suggestions and
illuminating discussions, as well as for his constant encouragement.
A.T. acknowledges the project no. 121720 of the Academy of Finland.




\begin{thebibliography}{99}
\bibitem{VSR}
  A.~G.~Cohen and S.~L.~Glashow,
  Phys.\ Rev.\ Lett.\  {\bf 97}, 021601 (2006).

\bibitem{CG}
  S.~R.~Coleman and S.~L.~Glashow,
  Phys.\ Rev.\  D {\bf 59}, 116008 (1999).

\bibitem{CK}
  D.~Colladay and V.~A.~Kostelecky,
  Phys.\ Rev.\  D {\bf 58}, 116002 (1998).




  \bibitem{monographs}

V. Chari and A. Pressley, \textit{A Guide to Quantum Groups}
(Cambridge University Press, Cambridge, 1994);\\
%
S. Majid, \textit{Foundations of Quantum Group Theory} (Cambridge
University Press, Cambridge, 1995);\\
%
M. Chaichian and A. Demichev, \textit{Introduction to Quantum
Groups} (World Scientific, Singapore, 1996).


  \bibitem{Drinfeld83}
  V. G. Drinfel'd,
  Sov. Math. Dokl. {\bf 27}, 68 (1983).
\bibitem{LAG}
L. \'Alvarez-Gaum\'e, J. L. F. Barb\'on and R. Zwicky, JHEP {\bf
0105}, 057 (2001).


\bibitem{CKNT}
M. Chaichian, P. Kulish, K. Nishijima and A. Tureanu, Phys. Lett. B
{\bf 604}, 98 (2004).

\bibitem{CPrT}
M. Chaichian, P. Pre\v{s}najder and A. Tureanu, Phys. Rev. Lett.
{\bf 94}, 151602 (2005).

\bibitem{LW}
J. Lukierski and M. Woronowicz, Phys. Lett. B {\bf 633}, 116 (2006).

\bibitem{CKTZZ}
M. Chaichian, P. P. Kulish, A. Tureanu, R. B. Zhang and Xiao Zhang,
J. Math. Phys. {\bf 49}, 042302 (2008).

\bibitem{CNST}
M. Chaichian, K. Nishijima, T. Salminen and A. Tureanu, JHEP {\bf
06}, 078 (2008).

\bibitem{DFR}
S. Doplicher, K. Fredenhagen and J. E. Roberts,
 Phys. Lett. B {\bf 331}, 39 (1994);
  Commun.\ Math.\ Phys.\  {\bf 172}, 187 (1995)



\bibitem{AAS}
F. Ardalan, H. Arfaei and M. M. Sheikh-Jabbari, JHEP {\bf 9902}, 016
(1999).


\bibitem{SW}
N. Seiberg and E. Witten, JHEP {\bf  9909}, 032  (1999).



\bibitem{AGM}
  O.~Aharony, J.~Gomis and T.~Mehen,
  JHEP {\bf 0009}, 023 (2000).


\bibitem{GM}
  J.~Gomis and T.~Mehen,
  Nucl.\ Phys.\  B {\bf 591}, 265 (2000).


\bibitem{CDP}
M. Chaichian, A. Demichev and P. Pre\v{s}najder, Nucl.\ Phys.\  B
{\bf 567}, 360 (2000)
  [arXiv:hep-th/9812180].



\bibitem{CH}
M. Chaichian and R. Hagedorn, {\it Symmetries in Quantum Mechanics:
From Angular Momentum to Supersymmetry}, IOP Publishing, 1998.



\bibitem{progress}
M. M. Sheikh-Jabbari and A. Tureanu (work in progress).

\bibitem{NCQFT-review}
  R.~J.~Szabo,
  Phys.\ Rept.\  {\bf 378}, 207 (2003).

  \bibitem{CPT}
  M.~M.~Sheikh-Jabbari,
  Phys.\ Rev.\ Lett.\  {\bf 84}, 5265 (2000).



\bibitem{CNT}
M.~Chaichian, K.~Nishijima and A.~Tureanu,
  Phys.\ Lett.\  B {\bf 568}, 146 (2003).

\bibitem{spin-stat}
A.~Tureanu,
  Phys.\ Lett.\  B {\bf 638}, 296 (2006);
  Prog. Theor. Phys. Suppl. {\bf 171}, 34
  (2007).








\bibitem{Lamb-shift}
  M.~Chaichian, M.~M.~Sheikh-Jabbari and A.~Tureanu,
  Phys.\ Rev.\ Lett.\  {\bf 86}, 2716 (2001).

\bibitem{Kostelecky}
  S.~M.~Carroll, J.~A.~Harvey, V.~A.~Kostelecky, C.~D.~Lane and T.~Okamoto,
  Phys.\ Rev.\ Lett.\  {\bf 87}, 141601 (2001).







\end{thebibliography}
\end{document}